\documentclass[prl,twocolumn,floatfix,dvipdfmx]{revtex4}
\usepackage{color}
\usepackage{graphicx}
\usepackage{bm}
\usepackage{amsmath}
\usepackage{indentfirst}
\usepackage{times}
\usepackage{latexsym}
\usepackage{txfonts,wasysym}

\definecolor{purple}{RGB}{223, 0, 255}
\begin{document}

\title{First-principles study on cubic pyrochlore iridates Y$_2$Ir$_2$O$_7$ and Pr$_2$Ir$_2$O$_7$}
\author{Fumiyuki Ishii$^{\rm 1}$\footnote{fishii@mail.kanazawa-u.ac.jp}, Yo Pierre Mizuta$^{\rm 2}$, Takehiro Kato$^{\rm 2}$, Taisuke Ozaki$^{\rm 3}$, Hongming Weng$^{\rm 4}$, and Shigeki Onoda$^{\rm 5,6}$}

\affiliation{
$^{\rm 1}$
Faculty of Mathematics and Physics,~Institute of Science and Engineering,~Kanazawa University,~Kanazawa,~920-1192,~Japan}
\affiliation{
$^{\rm 2}$ Graduate School of Natural Science and Technology, Kanazawa University, Kanazawa, 920-1192, Japan}
\affiliation{
$^{\rm 3}$ Institute for Solid State Physics, The University of Tokyo, Kashiwa 277-8581, Japan
}
\affiliation{$^{\rm 4}$Beijing National Laboratory for Condensed Matter
  Physics, and Institute of Physics, Chinese Academy of Sciences,
  Beijing 100190, China}
  \affiliation{
$^{\rm 5}$ Condensed Matter Theory Laboratory, RIKEN, Wako 351-0198, Japan}
\affiliation{
$^{\rm 6}$ RIKEN Center for Emergent Matter Science, Wako 351-0198, Japan
}

\date{\today}

\begin{abstract}
Fully relativistic first-principles electronic structure calculations based on a noncollinear local spin density approximation (LSDA) are performed for pyrochlore iridates Y$_2$Ir$_2$O$_7$ and Pr$_2$Ir$_2$O$_7$.
The all-in, all-out antiferromagnetic (AF) order is stablized by the on-site Coulomb repulsion $U>U_c$ in the LSDA+$U$ scheme, with $U_c\sim1.1$~eV and 1.3~eV for Y$_2$Ir$_2$O$_7$ and Pr$_2$Ir$_2$O$_7$, respectively. AF semimetals with and without Weyl points and then a topologically trivial AF insulator successively appear with further increasing $U$. For $U=1.3$~eV,  Y$_2$Ir$_2$O$_7$ is a topologically trivial narrow-gap AF insulator having an ordered local magnetic moment $\sim0.5\mu_B$/Ir, while Pr$_2$Ir$_2$O$_7$ is barely a paramagnetic semimetal with electron and hole concentrations of $0.016$/Ir, in overall agreements with experiments.  With decreasing oxygen position parameter $x$ describing the trigonal compression of IrO$_6$ octahedra, Pr$_2$Ir$_2$O$_7$ is driven through a non-Fermi-liquid semimetal having only an isolated Fermi point of $\Gamma_8^+$, showing a quadratic band touching, to a $Z_2$ topological insulator.  
\end{abstract}

\maketitle

Pyrochlore iridates $A_2$Ir$_2$O$_7$~\cite{Yanagishima_JPSJ2001} have attracted great interest for experimental observations of an itinerant chiral spin liquid~\cite{Machida_Nature2010} below 1.4~K 
and a possible underscreened Kondo effect around 20~K for $A$=Pr~\cite{Nakatsuji_PRL2006}, 
a finite-temperature transition from a semimetal to an antiferromagnetic (AF) insulator for $A$=Nd, Sm, and Eu and a Mott insulator for a shorter ionic radius of $A=$ Gd, $\cdots$, Lu and Y~\cite{Matsuhira_JPSJ2011,Taira_JPC2001,Disseler_PRB2012v2}. 
Ir $5d$ electrons are also advantageous for achieving topologically nontrivial states with and without electron correlations~\cite{Shitade_PRL2009}, because of a large relativistic spin-orbit coupling comparable to an electron transfer and the onsite Coulomb interaction $U$, as experimentally verified in Sr$_2$IrO$_4$~\cite{Kim_Science2009}. 
Theoretical proposals include a $Z_2$ topological band insulator (TBI)~\cite{Fu_Kane_PRB2007,Hasan_Kane_RMP2010,Pesin_NatPhys2010,Yang_Kim_PRB2010,Kurita_JPSJ2011} and Mott insulator~\cite{Pesin_NatPhys2010}, a Weyl semimetal and an axion insulator~\cite{Wan_PRB2011,Witczak-Krempa_PRB2013} accompanied by all-in, all-out magnetic order~\cite{Tomiyasu_JPSJ2012,Sagayama_PRB2013}, a rhombohedral $Z_2$ TBI under the $[111]$ pressure~\cite{Xiang_PRB2012}, and a non-Fermi-liquid semimetal~\cite{Moon:2013} associated with a quadratic band touching of electronic dispersions at the Fermi level~\cite{Yang_Kim_PRB2010,Yang_PRB2011}. 
However, understandings of the underlying detailed electronic structures from the first principles remain open to date.

One of the obstacles is that reliably treating localized rare-earth $4f$ electrons at the $A$ site is unfeasible in first-principles calculations. Calculations on Y$_2$Ir$_2$O$_7$ using crystal parameters for $A=$ Pr, Nd, Sm, and Eu on the less correlated side may yield a discrepancy, because the ionic radius and thus the lattice constant $a$ is longer by 1-3$\%$ in these materials than in Y$_2$Ir$_2$O$_7$. 
It is also crucial to understand the dependence of the electronic structure on the only two crystal structure parameters allowed for $A_2$Ir$_2$O$_7$, i.e., the lattice constant $a$ and the oxygen position parameter $x$~\cite{footnote_A2Ir2O7}, which vary with changing $A$ site elements.  
It has been argued that even in the least correlated material Pr$_2$Ir$_2$O$_7$, the all-in, all-out AF metal (AFM) is stable within the local spin-density approximation (LSDA)  based on the linear muffin-tin orbital method (LMTO) even for $U=0$~\cite{Wan_PRB2011}. This sharply contrasts to the experimental finding that Pr$_2$Ir$_2$O$_7$ is a paramagnetic semimetal (PSM)~\cite{Machida_Nature2010,Nakatsuji_PRL2006,Matsuhira_JPSJ2011,Taira_JPC2001}. 

\begin{figure*}[!htb!]
\begin{center}
\includegraphics[width=16cm]{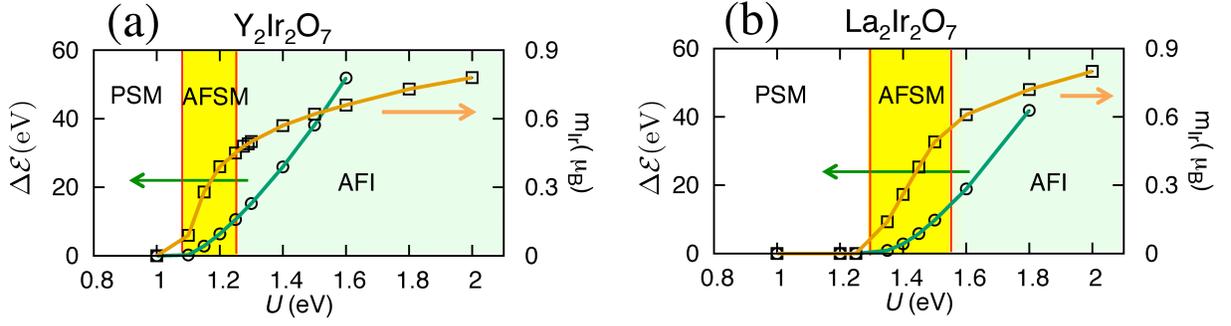}
\caption{(color online) 
Phase diagrams of (a) Y$_2$Ir$_2$O$_7$ and (b) Pr$_2$Ir$_2$O$_7$ as functions of $U$, obtained with the experimentally observed crystal parameters; the total energy difference ${\mit\Delta}\mathcal{E}$ between paramagnetic semimetal (PSM) and AF semimetal (AFSM) or insulator (AFI) in the left axis, and the magnitude $m_{\mathrm{Ir}}$ of the ordered local magnetic moment per Ir site in the right axes.}
\label{phase_U}
\end{center}
\end{figure*}

\begin{figure*}[!htb!]
\includegraphics[width=18cm]{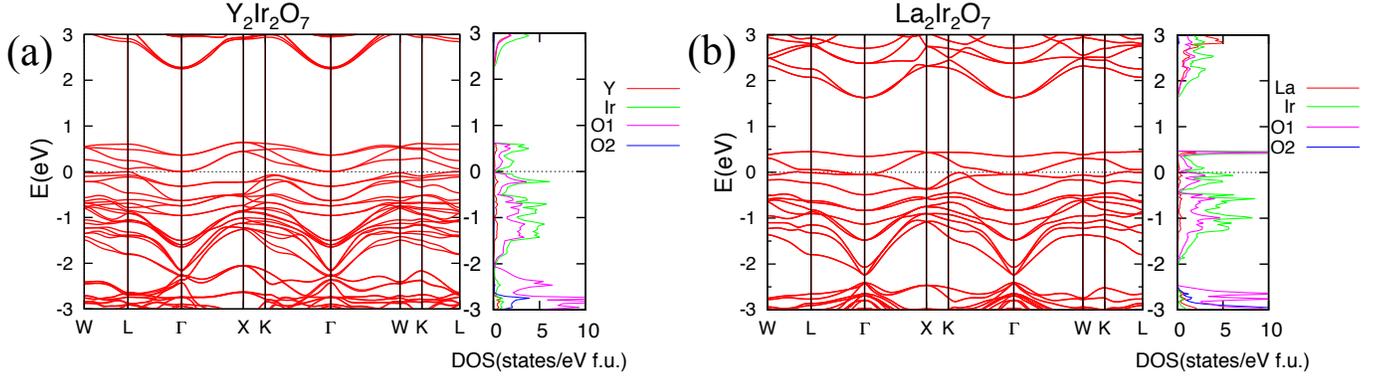}
\caption{(color online) Electronic band dispersions (left) and atom projected densities of states (right) of
(a) the all-in, all-out AFI for Y$_2$Ir$_2$O$_7$ and of
(b) the PSM for La$_2$Ir$_2$O$_7$ with the crystal parameters of Pr$_2$Ir$_2$O$_7$, 
obtained by the LSDA+$U$ ($U=1.3$~eV) and the LSDA ($U=0$), respectively. Red, green, magenta and blue curves represent Y or La, Ir, O1 ($48f$), and O2 ($8b)$ states, respectively. The Fermi level is denoted by grey dashed lines. }
\label{bands_LIO_YIO}
\end{figure*}

In this Letter, we provide extensive fully relativistic LSDA and LSDA+$U$ electronic structure calculations on Y$_2$Ir$_2$O$_7$ ($x$=0.335, $a=$10.106~\AA)~\cite{Shapiro_PRB2012} and on La$_2$Ir$_2$O$_7$ with varying crystal parameters around the experimentally observed values ($x$=0.330, $a$=10.400~\AA)~\cite{Millican2007} for Pr$_2$Ir$_2$O$_7$. A transition from a PSM to an all-in, all-out AF semimetal (AFSM) occurs at $U=U_c$ with $U_c\sim1.1$~eV and 1.3~eV for Y$_2$Ir$_2$O$_7$ and Pr$_2$Ir$_2$O$_7$, respectively (Fig.~\ref{phase_U}), in contrast to the previous LMTO results mentioned above~\cite{Wan_PRB2011}. 24 Weyl points~\cite{Wan_PRB2011} are observed at $U=1.15$-1.2~eV and 1.35-1.5~eV for Y$_2$Ir$_2$O$_7$ and Pr$_2$Ir$_2$O$_7$, respectively.
With further increasing $U$, the AFSM turns into a topologically trivial AF insulator (AFI). The AFI state of Y$_2$Ir$_2$O$_7$ with $U=1.3$~eV has an ordered local magnetic moment $m_{\mathrm{Ir}}\sim0.5\mu_B$/Ir, as shown in Fig.~\ref{phase_U}~(b), in reasonable agreement with the neutron-scattering experiments.~\cite{Shapiro_PRB2012} (The orbital and spin components are given by 0.29$\mu_B$ and 0.21$\mu_B$, respectively.) This value of $U$ is smaller than $U=2.0$~eV~\cite{Jin_PRB2009} in Sr$_2$IrO$_4$, probably because the localized nature of Ir $5d$ electrons is pronounced by the layered structure in Sr$_2$IrO$_4$. With the same value of $U\sim1.3$~eV, Pr$_2$Ir$_2$O$_7$ is located on the verge of the PM phase. (See Fig.~\ref{phase_U}~(a).) 
The PSM state for Pr$_2$Ir$_2$O$_7$ has a cuboidal electron Fermi surface centered at the $\Gamma$ point and toric hole Fermi surfaces around the $L$ points. With decreasing $x$ and thus increasing Ir-O-Ir bond angle from  $121^\circ$ ($x=0.35$) to $148^\circ$ ($x=0.30$), we observe a transition to a Fermi-point semimetal hosting a quadratic band touching of two doubly degenerate bands which form $\Gamma_8^+$ at the Fermi level. A further decrease in $x$ drives a level crossing of $\Gamma_8^+$ and $\Gamma_6^+$, resulting in a $Z_2$ topological insulator for $x\sim0.31$.

The calculations have been performed by means of a fully-relativistic non-collinear spin density functional theory within the LSDA and LSDA+$U$ methods implemented in OpenMX code~\cite{OpenMX}, 
which is based on the norm-conserving pseudopotential and a linear combination of multiple pseudo-atomic orbitals. (For computational details, see Supplemental Materials.) We use (9,9,9) to (18,18,18) uniform $k$-point mesh. 
The pseudopotentials and pseudo-atomic orbitals are severely checked by comparison with the all-electron full-potential linearized augmented plane-wave (FLAPW) method~\cite{HiLAPW}, 
so that our fully relativistic scheme is well compared to the FLAPW results in the LDA with the spin-orbit interaction being treated in the second variation scheme. (See Fig.~S1 in Supplemental Materials.)
The total-energy calculation using OpenMX reveals a structural stability around  $a=10.4~{\mathrm \AA}$ and $x=0.33$ for La$_2$Ir$_2$O$_7$ (see Supplemental Materials), which are close to the experimental observations for Pr$_2$Ir$_2$O$_7$ and justifies the La replacement as long as the low-temperature rare-earth magnetism is irrelevant above 1.5~K\cite{Machida_Nature2010}. Thus, we henceforth replace Pr with La for calculations on Pr$_2$Ir$_2$O$_7$.
We assume that the translational invariance is fully preserved, while the ordered magnetic moment directions have been fully relaxed during the calcuations without adding any energy barrier and converged to the all-in, all-out AF structure in all the magnetically ordered cases shown below. 

Figure \ref{bands_LIO_YIO} shows the band structure and atom projected density of states (DOS) of (a) a narrow-gap AFI obtained for Y$_2$Ir$_2$O$_7$ with the LSDA+$U$ ($U=1.3$~eV) and (b) a PSM obtained for Pr$_2$Ir$_2$O$_7$ with the LSDA ($U=0$). The Fermi level has been chosen to be $E_{\mathrm F}=0$.
In both cases, the bands in $E\in[-2, 0.5]$~eV shown in the atom projected DOS (Fig.\ref{bands_LIO_YIO}) are mainly composed of anti-bonding states of Ir $5d$ and O1 ($48f$) $2p$ electrons in IrO$_6$ octahedra. The energy windows $E\in[0.5,2.0]$~eV for Y$_2$Ir$_2$O$_7$ and $E\in[0.5,1.5]$~eV for Pr$_2$Ir$_2$O$_7$ are given by a crystal-field gap due to the IrO$_6$ octahedral coordination. 

\begin{figure}[!b]
\begin{center}
\includegraphics[width=9.0cm]{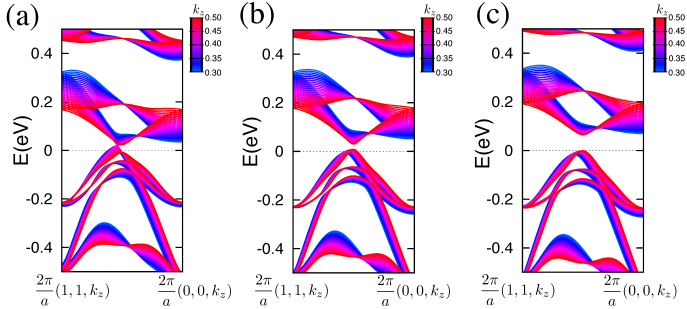}
\caption{(Color online) Electron band dispersions in the $\Gamma$-$X$-$L$ plane for Y$_2$Ir$_2$O$_7$, (a) in the AFSM with $U=1.2$~eV, (b) around the AFSM-AFI boundary at $U=1.25$~eV, and (c) in the AFI with $U=1.3$~eV.}
\label{weyl}
\end{center}
\end{figure}

In this AFI for Y$_2$Ir$_2$O$_7$, the charge gap is suppressed by a proximity to the AFWSM. Figure~\ref{weyl} shows an evolution of the electron dispersions around a Weyl point with increasing $U$ from 1.2 to 1.3~eV. Weyl points are confined to the $\Gamma$-$X$-$L$ and symmetry-related planes~\cite{Wan_PRB2011} by the magnetic space group symmetry $Fd\bar{3}m'$ if they exist. They indeed appear at $\bm{k}=\frac{2\pi}{a}$(0.44, 0.44, 0.35) and (0.46,0.46,0.41) for $U=1.15$~eV and 1.2~eV, respectively, and symmetry-related points, with their energy level being located about 35 and 24~meV higher than the Fermi level, leaving electron and hole Fermi surfaces associated with the concentration $\sim0.008/\mathrm{Ir}=1.3\times10^{20}$~/cm$^3$. The Weyl point shifts towards the $L$ point with increasing $U$ and eventually disappear at $U\sim1.25$~eV. This leaves an topologically trivial AFSM with similar electron and hole Fermi pockets. Then, an indirect gap $\sim5$~meV opens between the valence band top at the $L$ point and the conduction band bottom at the $\Gamma$ point for $U=1.3$~eV. It is likely that disorder effects localize thermally excited dilute carriers in this narrow-gap AFI and inelastic scattering events beyond the LSDA+$U$ shcheme give rise to the variable range hopping, as experimentally observed in Y$_2$Ir$_2$O$_7$~\cite{Liu_SSC2013}, as well as in Nd$_2$Ir$_2$O$_7$~\cite{Matsuhira_JPSJ2013} and Eu$_2$Ir$_2$O$_7$\cite{Ishikawa_PRB2012}.

\begin{figure}[!b]
\begin{center}
\includegraphics[width=6cm]{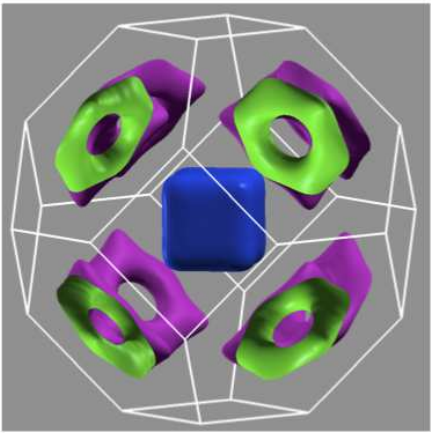}
\caption{(color online) Fermi surfaces of La$_2$Ir$_2$O$_7$ for the crystal structure of Pr$_2$Ir$_2$O$_7$. 
}
\label{FermiSurface}
\end{center}
\end{figure}

The PSM for Pr$_2$Ir$_2$O$_7$ contains a twofold degenerate cuboidal electron Fermi surface around the $\Gamma$ point and twofold degenerate hole pockets around the $L$ points, as shown in Fig.~\ref{FermiSurface}. From the Fermi volumes, both the electron and hole carrier numbers are given by $n_e=n_h=n=0.016/\mathrm{Ir}=2.22\times10^{20}$~/cm$^3$ for Pr$_2$Ir$_2$O$_7$. From the Hall coefficient $R_H=-2.4\times10^{-2}$~cm$^3$/C measured in Pr$_2$Ir$_2$O$_7$~\cite{Nakatsuji_PRL2006}, the ratio of the electron and hole mobilities $\mu_e$ and $\mu_h$ is estimated to be a rather large value, $\mu_e/\mu_h=\frac{1+neR_H}{1-neR_H}=13$, through $R_H=\frac{(n_e\mu_e^2-n_h\mu_h^2)}{e(n_e\mu_e+n_h\mu_h)^2}$ with the electron charge $e$. 

\begin{figure}[!h]
\begin{center}
\includegraphics[width=9cm]{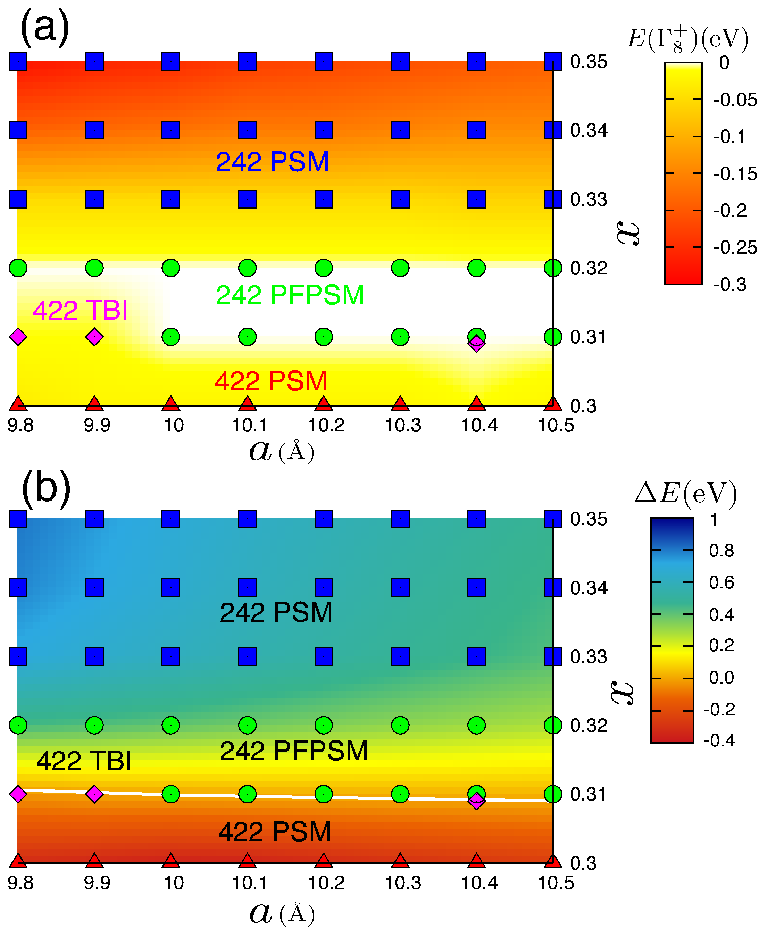}
\caption{(color online) Phase diagrams of La$_2$Ir$_2$O$_7$: (a) the energy level $E(\Gamma_8^+)-E_{\mathrm{F}}$
and (b) the level difference ${\mit\Delta}E=E(\Gamma_8^+)-E(\Gamma_6^+)$, which changes the sign at the white solid line.
$\textcolor{blue}{\blacksquare}$: a 242 PSM, 
$\textcolor{green}{\CIRCLE}$: a 242 paramagnetic Fermi-point semimetal (PFPSM), $\textcolor{purple}{\Diamondblack}$: a 422 $Z_2$ TBI, and $\textcolor{red}{\blacktriangle}$: a 422 PSM.
The numbers 242 and 422 in the above represent those of the degeneracies of the three levels at the $\Gamma$ point around $E_{\rm F}$ in the ascending order in energy.}
\label{phasediagram}
\end{center}
\end{figure}

\begin{figure}[!h]
\begin{center}
\includegraphics[width=8.0cm]{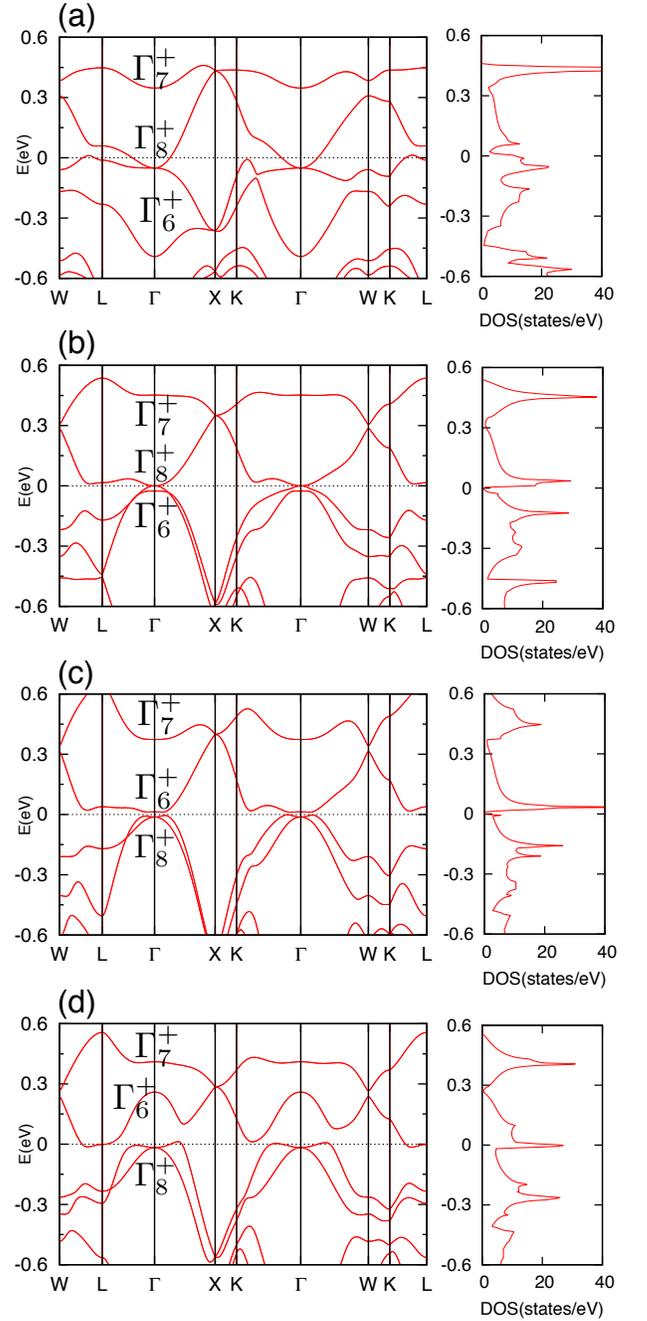}
\caption{(color online) The $(a,x)$ dependence of the band structure of La$_2$Ir$_2$O$_7$: (a) ($a=10.4$~\AA, $x=0.33$) the same structure as Pr$_2$Ir$_2$O$_7$, (b) ($a=10.4$~\AA, $x=0.31$), (c) ($a=9.8$~\AA, $x=0.31$), and (d) ($a=10.4$~\AA, $x=0.30$). }
\label{bands}
\end{center}
\end{figure}

Now we study the dependence of the electronic structures on $a$ and $x$ within a PSM or a band insulator, as $(a,x)$ in Pr$_2$Ir$_2$O$_7$ depends on the temperature, a hydrostatic pressure, and possibly the sample quality. Figure~\ref{phasediagram} shows the phase diagram of La$_2$Ir$_2$O$_7$ obtained with $U=0$. It contains four phases; (a) a PSM as obtained for Pr$_2$Ir$_2$O$_7$, (b) a paramagnetic Fermi-point semimetal (PFPSM) of two doubly degenerate quadratic bands touching at the $\Gamma$ point at $E_{\mathrm{F}}$ from above and below, (c) a $Z_2$ TBI, and (d) another PSM with even smaller electron and hole Fermi pockets than in Pr$_2$Ir$_2$O$_7$. This evolution of the electronic structure is attributed to a change in the single-orbital/Ir Kramers degenerate states closest to $E_{\mathrm{F}}$, which form a closed manifold consisting of four doubly degenerate bands, as shown in Figs.~\ref{bands}~(a) to (d), respectively.
In particular, the key role is played by an energy level crossing of a twofold degenerate $\Gamma_6^+$ and a fourfold degenerate $\Gamma_8^+$, as we show below. (The degeneracy of the latter is protected by the cubic and time-reversal symmetry, while that of the former is only by the time-reversal.)

When $x>x_c\sim0.31$, namely, in the upper part than a white solid line in the phase diagram (Fig.~\ref{phasediagram}~(b)), the energy levels at the $\Gamma$ point satisfy the relation $E(\Gamma_6^+)<E(\Gamma_8^+)<E(\Gamma_7^+)$, as shown in Fig.~\ref{bands}(a). With decreasing $x$, $E(\Gamma_8^+)$ becomes shallow below $E_{\mathrm{F}}$, as seen from the color map of Fig.~\ref{phasediagram}~(a). Accordingly, both electron and hole Fermi volumes shown for $a=10.40~\mathrm{\AA}$ and $x=0.33$ in Fig.~\ref{FermiSurface} gradually decrease. This slight increase of $E(\Gamma_8^+)$ appears to be due to a rapid increase of $E(\Gamma_6^+)$ approaching $E(\Gamma_8^+)$ and $E_{\mathrm{F}}$, as shown in the color map of Fig.~\ref{phasediagram}~(b). Then, around $x\sim0.32$, the electron Fermi surface eventually shrinks into a single point, while the hole Fermi surfaces disappear, yielding a PSM hosting a single Fermi point with $E(\Gamma_8^+)=E_{\mathrm{F}}$, as shown in Fig.~\ref{bands}~(b). This paramagnetic Fermi-point semimetal (PFPSM) appears in a white region in Fig.~\ref{phasediagram}~(a). (To be precise, we obtained this solution for $9.8~{\mathrm \AA} \le a \le 10.5~{\mathrm \AA}$ in the case of $x=0.32$ and for $10.0~{\mathrm \AA} \le a \le 10.5~{\mathrm \AA}$ in the case of $x=0.31$, as marked by green points in Fig.~\ref{phasediagram}.) 

On the other hand, with decreasing $x$ below $x_c$, $E(\Gamma_6^+)$ becomes larger than $E(\Gamma_8^+)$ in the lower part of the phase diagram than the white solid line (Fig.~\ref{phasediagram}~(b)). This yields a narrow-gap insulator around $x\sim0.31$. (To be precise, the insulating solution is found for $x=0.31$ in the case of $a=9.8$ and 9.9~\AA \ 
and for $x=0.309$ in the case of $a=10.4$~\AA, as shown in Fig.~\ref{bands}(c).) This is actually a strong $Z_2$ TBI with the $Z_2$ topological indices~\cite{Fu_Kane_PRB2007} $(1;0,0,0)$, which has been confirmed by directly calculating the parity eigenvalues of all the occupied states at the time-reversal invariant momenta. The same topological indices can also be obtained within the single-orbital manifold per Ir near $E_{\mathrm{F}}$. Within this manifold, the occupied states at the eight time-reversal invariant momenta take the following irreducible representations; the fourfold degenerate $\Gamma_8^+$, fourfold degenerate $X_5^++X_5^-$ at $X:(\frac{\pi}{a},0,0)$ and the two symmetry-related points, doubly degenerate $(L_4^++L_5^+)$ and $L_6^+$ at a single $L$ point, e.g., $(\frac{\pi}{a}, \frac{\pi}{a}, \frac{\pi}{a})$, and $(L_4^-+L_5^-)$ and $L_6^-$ at the three symmetry-related points.  Note that the $Z_2$ topological indices do not altered by any level crossing, for instance, of $E(\Gamma_6^+)$ and $E(\Gamma_8^+)$, observed in the above phase diagram. Actually, the $Z_2$ TBI is realized by simply gapping out the band dispersions around the Fermi level, which certainly requires $E(\Gamma_8^+)<E(\Gamma_6^+),E(\Gamma_7^+)$. 
With further decrease in $x$ down to 0.30, we observe that bands cross the Fermi level, leaving small electron and hole Fermi pockets near the $\Gamma$ and $L$ points, respectively, for $9.8~{\mathrm \AA} \le a \le 10.5~{\mathrm \AA}$, as shown in Fig \ref{bands}(d). Note that the $Z_2$ TBI phase and the PFPSM phase may expand more if we employ a method which may improve the bandgap in LDA calculations, such as modified Becke-Johnson exchange potential\cite{TB-mBJ_PRL2009}, hybrid functionals(e.g., HSE\cite{HSE_JCP2003}), and GW methods\cite{SCF-GW_PRL2004}.

The PSM solution obtained for the crystal parameters of Pr$_2$Ir$_2$O$_7$ is clearly compatible with the experimental finding on the ordinary Hall constant~\cite{Nakatsuji_PRL2006}, as we already discussed, if we assume that the mobility of the holes is an order of magnitude smaller than that of the electrons. In this scenario, Shubnikov-de Haas oscillations with a single Lifshitz-Konsevich oscillatory component, experimentally observed in Pr$_2$Ir$_2$O$_7$~\cite{Balicas_PRL2011}, should be ascribed to the electron-like carriers. 
It will be possible to reach the quadratic band touching PFPSM, if a compound with smaller $x$ is found.  This state can be experimentally verified by a $\sqrt{H}$-dependent Hall resistivity~\cite{Moon:2013} and an angle-resolved photoemission spectroscopy. 
The bulk energy gap in our cubic $Z_2$ topological insulator phase can be as large as 20~meV, and is experimentally observable, although it is required to find a material with even smaller $x$. 

When we almost finished the work, we noticed recent preprints~\cite{Shinaoka_arxiv2015,Zhang_arxiv2015} on the LSDA+DMFT calculations for $A_2$Ir$_2$O$_7$, which show a direct first-order phase transition between the PM and an AFI in the space of $U$ and the temperature. The stability of the paramagnetic state up to a moderate value of $U$ is consistent with our results. However, $U_c$ for the AF order is much larger than in our current results, which washes out the AFSM with or without Weyl points in the weak-to-intermediate range of $U$. The ordered local magnetic moment, $0.65\mu_B$~\cite{Shinaoka_arxiv2015} or $0.57\mu_B$~\cite{Zhang_arxiv2015}, for Y$_2$Ir$_2$O$_7$ and the charge gap, 0.3~eV for Y$_2$Ir$_2$O$_7$~\cite{Shinaoka_arxiv2015} or 0.4-0.5~eV for Nd$_2$Ir$_2$O$_7$~\cite{Zhang_arxiv2015}, are also overestimated in comparison with our current results which agree with experiments. Note also that a recent angle-resolved photoemission spectroscopy measurement in the low-temperature all-in, all-out ordered state of Nd$_2$Ir$_2$O$_7$~\cite{Tomiyasu_JPSJ2012} has revealed half the symmetrized direct energy gap of 50~meV at the $L$ point~\cite{Nakayama_JPSmeeting}, which is compatible with our results (Fig.3(c)) but not the LSDA+DMFT results~\cite{Shinaoka_arxiv2015,Zhang_arxiv2015}.

The authors thank S. Nakatsuji, K. Matsuhira, and Y.-B. Kim for discussions. 
The work was partially supported by Grants-in-Aid on Scientific Research under Grant Nos. 24740253, 25104714, 25790007, 15H01015 and 15H03692 from Japan Society for the Promotion of Science, by the RIKEN iTHES Project, and by the MEXT HPCI Strategic Program. 
S.O. acknowledges the hospitality of the Aspen Center for Physics, which was supported by the National Science Foundation under Grant No. PHYS-1066293, and the hospitality of Nordita, Nordic Institute of Theoretical Physics, during his stays at the end of the work. 
Computations in this research were performed using RIKEN Integrated Cluster of Clusters, supercomputers at ISSP, University of Tokyo, and supercomputers at RIIT, Kyushu University.


\begin{thebibliography}{36}
\bibitem[{foo()}]{footnote_A2Ir2O7}
\bibinfo{note}{$A_2$Ir$_2$O$_7$ ($A$: Y, rare-earth elements) compounds form
  the pyrochlore structure having the space group $Fd{\bar{3}m}$. $A^{3+}$ and
  Ir$^{4+}$ ions are located at the Wykoff positions $16d$
  $(\frac{1}{2},\frac{1}{2},\frac{1}{2})$ and $16c$ $(0,0,0)$, respectively,
  while O$^{-2}$ ions are located at the Wykoff positions $48f$
  $(x,\frac{1}{8},\frac{1}{8})$ (O1) and $8b$
  $(\frac{3}{8},\frac{3}{8},\frac{3}{8})$ (O2).}
\end{thebibliography}

\newpage
\widetext
\begin{center}
	\textbf{\large Supplemental Materials}
\end{center}
\setcounter{equation}{0}
\setcounter{figure}{0}
\setcounter{table}{0}
\setcounter{page}{1}
\makeatletter
\renewcommand{\theequation}{S\arabic{equation}}
\renewcommand{\thefigure}{S\arabic{figure}}
\renewcommand{\bibnumfmt}[1]{[S#1]}

\section{Computational Methods}
\subsection{OpenMX}
By using the OpenMX code\cite{OpenMX_S}, we have performed fully-relativistic first-principles band structure calculations based on the density functional theory (DFT) within the local density approximation (LDA)\cite{Ceperley1980_S}.
Fully relativistic pseudopotentials including spin-orbit coupling are generated by the Morrison-Bylander-Kleinman scheme\cite{Morrison1993_S}, which includes a partial core correction\cite{PCC_S}.
The wave functions were expanded in a linear combination of multiple pseudo-atomic orbitals (LCPAO) generated by a confinement scheme\cite{PAO1_S, PAO2_S}. 
Pseudoatomic orbitals are chosen to be 
La8.0-$s$3$p$3$d$1$f$1, Ir7.0-$s$3$p$3$d$2, and O5.0-$s$3$p$3$d$1.
The numbers after the chemical symbol (8.0, 7.0, 5.0) are the cutoff radii 
(bohr), while the latter parts of the expansion (e.g., $s$3$p$3$d$1$f$1 etc.) are the number of orbitals for the $s$, $p$, $d$ and $f$ characters.
We adopted the cutoff energy 250 Ry which was used in the calculation of matrix elements associated with the difference charge Coulomb potential and the exchange-correlation potential, and in solving Poisson's equation using fast Fourier transform (FFT).

\subsection{FLAPW}
We include spin-orbit coupling as the second variation after convergence of the scalar-relativistic self-consistent-field (SCF), i.e., we diagonalize the Hamiltonian matrix including spin-orbit interaction constructed from scalar-relativistic eigenstates and wavefunctions\cite{HiLAPWso1_S,HiLAPWso2_S}. 
We use  Perdew-Zunger expression for exchange and correlation energy\cite{PZ1981_S}. 
The ${\bf k}$-space integration is achieved by the improved tetrahedron method\cite{Blochl1994_S}. 
For the face centered cubic cell, a (4,4,4) mesh is used during 
the SCF loops with 8 $\bm{k}$ point in the irreducible Brillouin zone (IBZ). 
Muffin-tin spheres radii assumed are 1.0 \AA  \ for La, 1.1 \AA \ for Ir, and 0.8 \AA \ for O.
Core states of 1$s$ to 4$d^{10}$ for La, 5$p^{6}$ for Ir, 1$s^2$ for O 
fully relaxed by using the spherical part of the one-electron potential
 during the SCF iterations. The energy cutoff of plane waves used is 25 Ry. 
 The charge density and potential are expanded using the cutoff of 100 Ry. 

\section{Checking the pseudopotentials and the pseudo-atomic orbitals}
\begin{figure}[ht]
\begin{center}
\includegraphics[width=16cm,angle=0]{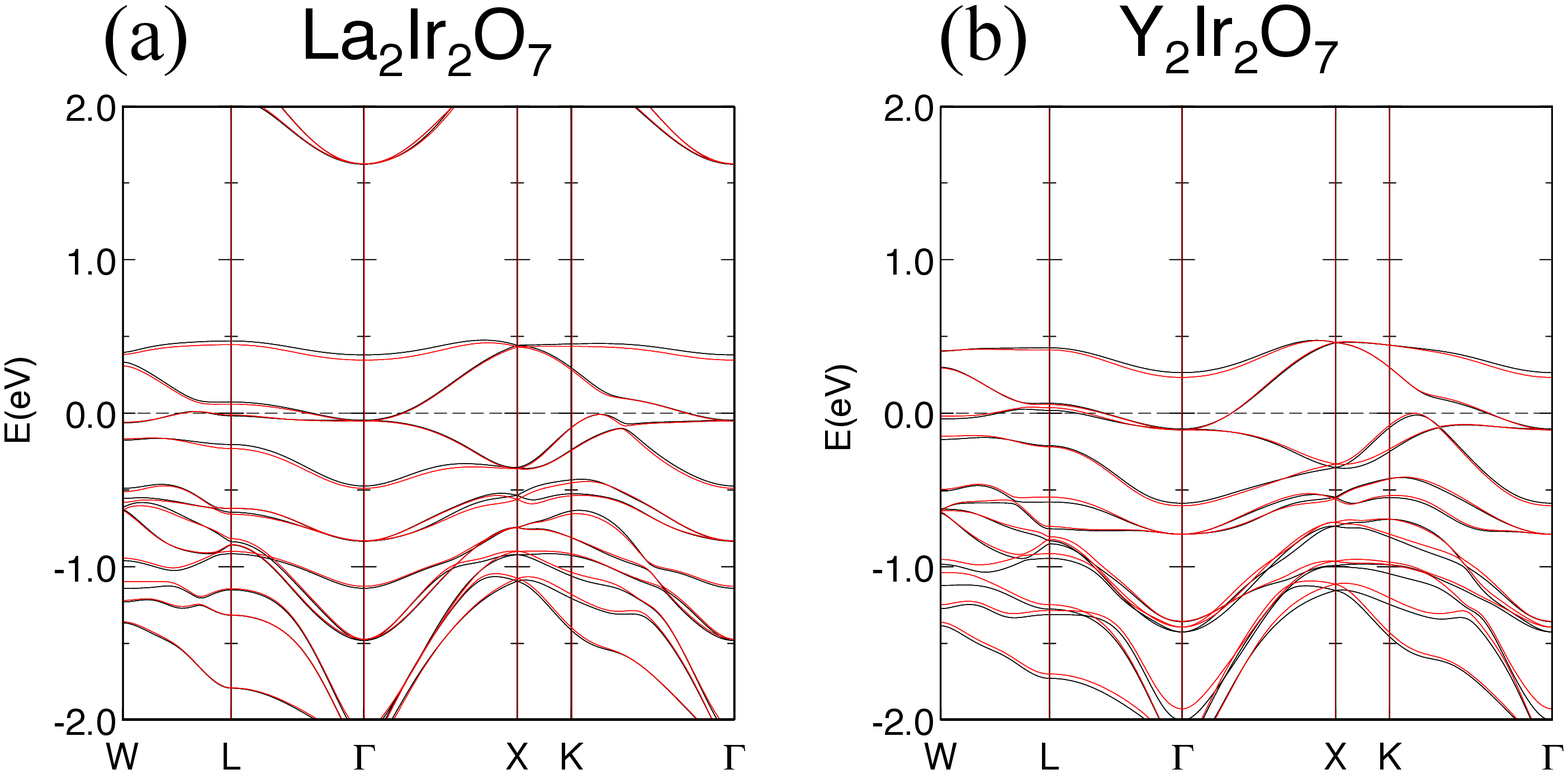}
\caption
{ 
Calculated band structure of (a) La$_2$Ir$_2$O$_7$ and  (b) Y$_2$Ir$_2$O$_7$.
Black lines are calculated by HiLAPW code. Red lines are calculated by OpenMX code.
}
\label{band_La_Y}
\end{center}
\end{figure}

In order to check the reliability of the pseudopotential, calculations were also done with the all-electron full-potential linearized augmented plane-wave (FLAPW) methods~\cite{HiLAPW_S, Ishii_JPSJ2000_S, Ishii_JPSJ2004_S} with local density approximation (LDA) including spin-orbit interaction. 
Figure \ref{band_La_Y} shows comparison of the band structures of FLAPW methods (HiLAPW code) and OpenMX code, for (a) La$_2$Ir$_2$O$_7$ with the crystal parameters of Pr$_2$Ir$_2$O$_7$ and (b) Y$_2$Ir$_2$O$_7$ with the relativistic spin-orbit interaction. 
The calculated band structures obtained with the OpenMX code show good agreements with those with the HiLAPW code. 
This gives a support for the reliability of the present pseudopotential calculations and 
convergence in pseudo-atomic orbitals. 

\section{Structural Stability}
\begin{figure}[h]
\begin{center}
\includegraphics[width=12cm]{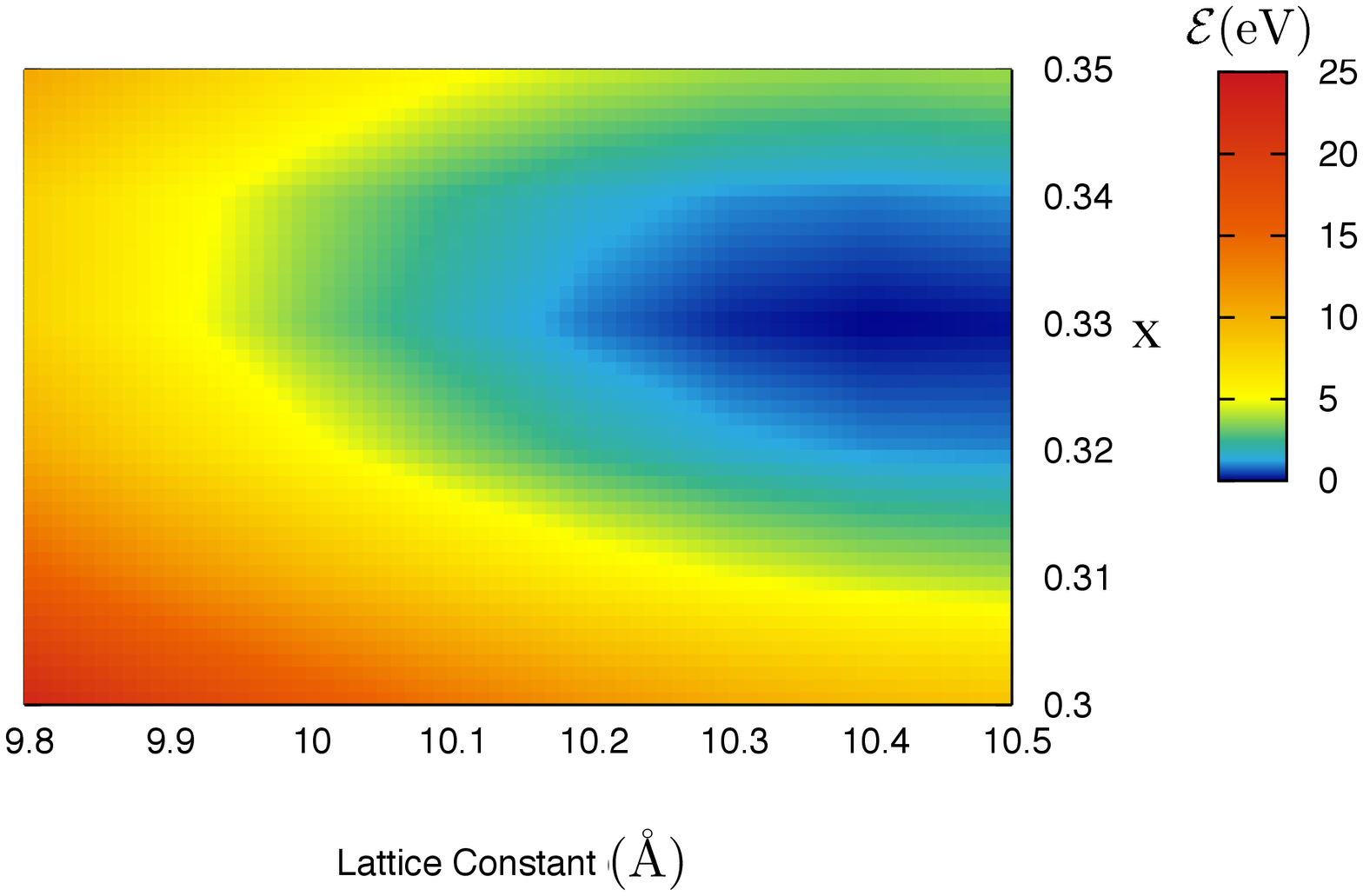}
\caption{
Total energy for La$_2$Ir$_2$O$_7$, the lattice constant $a$ and internal parameter $x$ for oxygen 48$f$ site  changes }
\label{fig6}
\end{center}
\end{figure}

Figure \ref{fig6} shows total-energy profile for La$_2$Ir$_2$O$_7$ in terms of lattice constant $a$ and
internal parameter $x$ for oxygen 48$f$ site. The calculated total-energy is lowest at 
the experimentally observed $a$ and $x$ for Pr$_2$Ir$_2$O$_7$. 

\end{document}